\documentclass[12pt]{article}       
\usepackage{dina4p}
\usepackage{epsfig}

\def\be{\begin{equation}}
\def\ee{\end{equation}}
\begin{document}
\begin{center}  
\begin{Large}
{\bf Future Measurements of Transversity\footnote{Talk presented by V.A.~Korotkov
at the International Workshop "Symmetries and Spin, Praha-Spin-2000", 
Prague, July 17-22, 2000} }
\end{Large}

\vspace*{+4mm}

\begin{large}
V.A.~Korotkov$^{a,b}$ and W.-D.~Nowak$^a$ 
\end{large}

\vspace*{+4mm}

$^a$ DESY Zeuthen, D-15735~Zeuthen, Germany \\
$^b$ IHEP, RU-142284 Protvino, Russia \\
\end{center}

\vspace*{+4mm}

{\bf Abstract:}
A review of envisaged future quark transversity measurements 
is presented.

\section{Introduction}
\label{sect1}
%

A quark of a given flavor in the nucleon is characterized by three 
twist-two quark distributions: the number density distribution
$q(x)$, the helicity distribution $\Delta q(x)$, and
the experimentally unknown transversity distribution $\delta q(x)$, 
which characterizes the distribution of the quark's transverse spin in a 
transversely polarized nucleon. 
For  non-relativistic quarks $\delta q(x) = \Delta q(x)$ can be expected,
but generally  both distribution functions are independent. 
Soffer's inequality for each quark flavor,
$2 |\delta q(x, Q^2)| \leq q(x, Q^2) + \Delta q(x, Q^2)$, 
restricts possible values of the transversity distributions.

The transversity distribution was first discussed by Ralston and Soper 
\cite{ralston} in doubly transverse polarized Drell-Yan (DY) scattering. Its 
measurement is one of the main goals of the spin program at RHIC. 
The transversity distributions $\delta q(x)$ are not accessible in
inclusive DIS, because they are chiral-odd and only occur
in combinations with other chiral-odd objects.
In semi-inclusive DIS of unpolarized leptons off transversely
polarized nucleons several methods have been proposed to access 
$\delta q(x)$ via specific single target-spin asymmetries.

\section{Measurement of $\delta q (x)$ in pp Collisions} 

\begin{figure}[htb]
\centering
\begin{minipage}[c]{10.0cm}
\centering
\epsfig{file=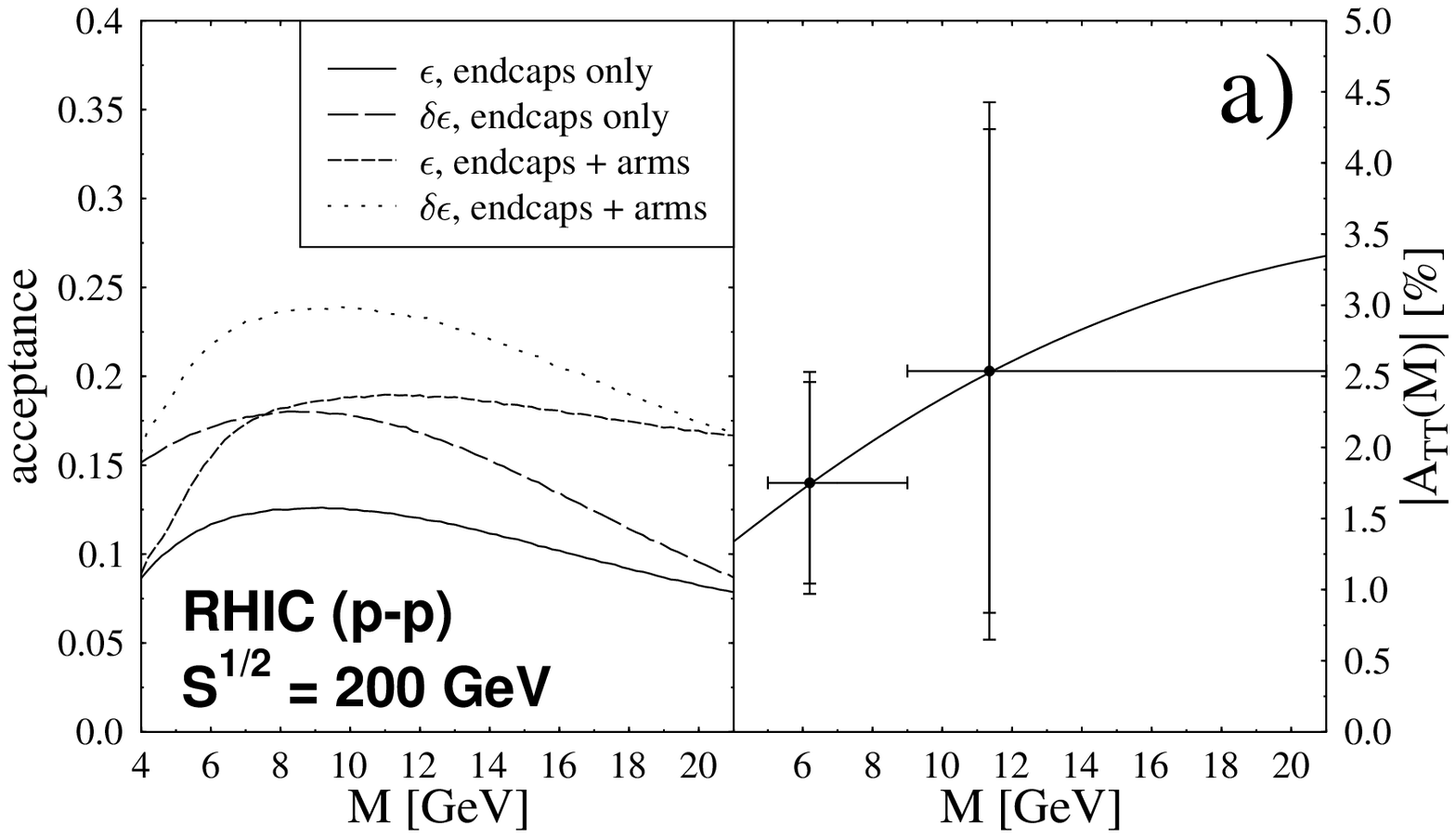,width=9.1cm}
\end{minipage}
\begin{minipage}[c]{6.5cm}
\centering
\epsfig{file=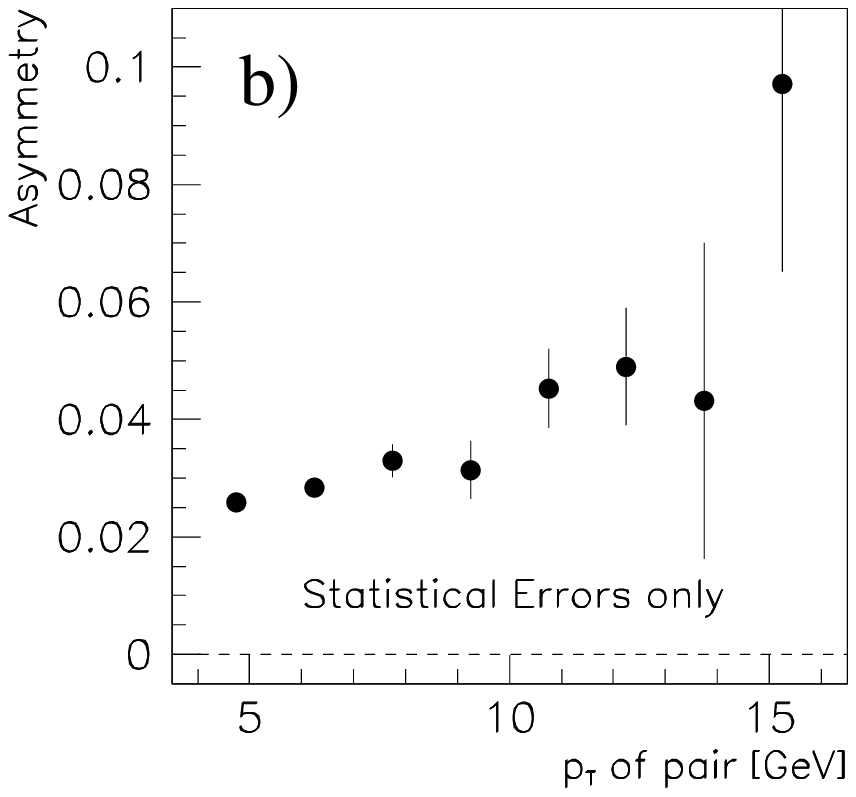,width=5.9cm}
\end{minipage}
\caption{\it Projections for measurements  with PHENIX
 ($\sqrt{s} = 200$~GeV) of a) Drell-Yan asymmetry 
$A_{TT}$ ($L = $ 320~pb$^{-1}$)~\cite{omartin}, and b) two-pion 
asymmetry ($L = $ 32~pb$^{-1}$)~\cite{matthias}. }
\label{ppasymm}
\end{figure}

An evaluation of the DY
asymmetry $A_{TT} \sim  \sum e_{i}^{2} \delta q_i (x) \delta \bar{q_i}(x)$ 
was carried out \cite{omartin} by assuming the saturation of Soffer's inequality 
for the transversity distributions. The maximum possible asymmetry at RHIC 
energies was estimated to be $A_{TT} = 1 \div 2 \%$ (see fig.\ref{ppasymm}a). 
At smaller energies, e.g. for a possible fixed-target 
experiment HERA-$\vec N$ \cite{heran} ($\sqrt{s} \simeq 40$ GeV), 
the asymmetry is expected to be higher.

A better sensitivity to $\delta q(x)$ is expected in
a measurement of two-meson correlations with the nucleon's transverse spin.
The interference effect between the
$s$- and $p$-waves of the two-meson system
allows the  quark's
polarization information to be  carried through $\vec k_+
\times \vec k_- \cdot  \vec S_\perp$ \cite{jaffe97a,tang}.
Here, $\vec k_+$ and $\vec k_-$ are the meson momenta, and $\vec S_\perp$ is
the proton spin vector.
The corresponding asymmetry depends on the unknown chiral-odd interference 
quark fragmentation function (FF), $\delta \hat{q}_{_I}(z)$. 
The function $\delta \hat{q}_{_I}(z)$
has a theoretical upper bound and could be measured in 
$e^+e^-  \rightarrow  (\pi^+\pi^-X) (\pi^+\pi^-X)$.
To estimate a possible level of the asymmetry at RHIC energies 
two assumptions were made~\cite{tang}: 
i) $\delta q (x, Q^2)$ saturates Soffer's inequality, and
ii) $\delta \hat{q}_I( z )$ saturates its upper bound.
This approach produces the {\it maximally possible} asymmetry. 
The projections of the asymmetry measurement with PHENIX at RHIC~\cite{matthias}
are shown in fig.~\ref{ppasymm}b.


\section{Measurements of $\delta q (x)$ in SIDIS} 

In semi-inclusive deep inelastic lepton scattering (SIDIS) off transversely 
polarized nucleons there exist several methods to access transversity 
distributions. One of
them, namely twist-3 pion production \cite{jaffe-ji93}, uses 
longitudinally polarized leptons and a double spin asymmetry is measured. 
The other methods do not require a polarized beam, they rely on 
polarimetry of the scattered transversely polarized quark:
i) measurement of the transverse polarization of $\Lambda$'s in the 
current fragmentation region \cite{jaffe96};
ii) observation of a correlation between the nucleon's transverse spin vector
 and the normal to the two-meson plane \cite{jaffe97a};
iii) observation of the Collins effect in quark fragmentation through 
 the measurement of pion single target-spin asymmetries \cite{collins}.    
HERMES data \cite{hermes} indicate that the polarized FF $H_1^{\perp q}(z)$,
responsible for the Collins effect, is quite sizeable.

\subsection{Future Measurements at HERMES}

The expected statistics for running at HERMES ($E = 27.5$~GeV) 
with a transversely polarized proton target will
consist of about seven millions reconstructed DIS events.
As average beam and target polarizations $P_B = 50 \%$ and $P_T = 75 \%$,
respectively, are used for the analysis. 
DIS events are defined as those satisfying the following set of kinematic 
cuts:
$Q^2  \ > 1 $~GeV$^2$, $W  \ > 2 $~GeV, 
$0.02 \ < \  x  \ < \ 0.7$, $y  \ < \ 0.85$. 
The following cuts were assumed for the kinematic variables of the pion:
$x_F > 0.$, $z > 0.1$, $P_{h \perp} > 0.05$ GeV. \\
The approximation $\delta q(x)  =  \Delta q(x)$ could be
used for the evaluation of the below given projections in view of the relatively 
low $Q^2$-values at HERMES.

\begin{figure}
\centering
\epsfig{file=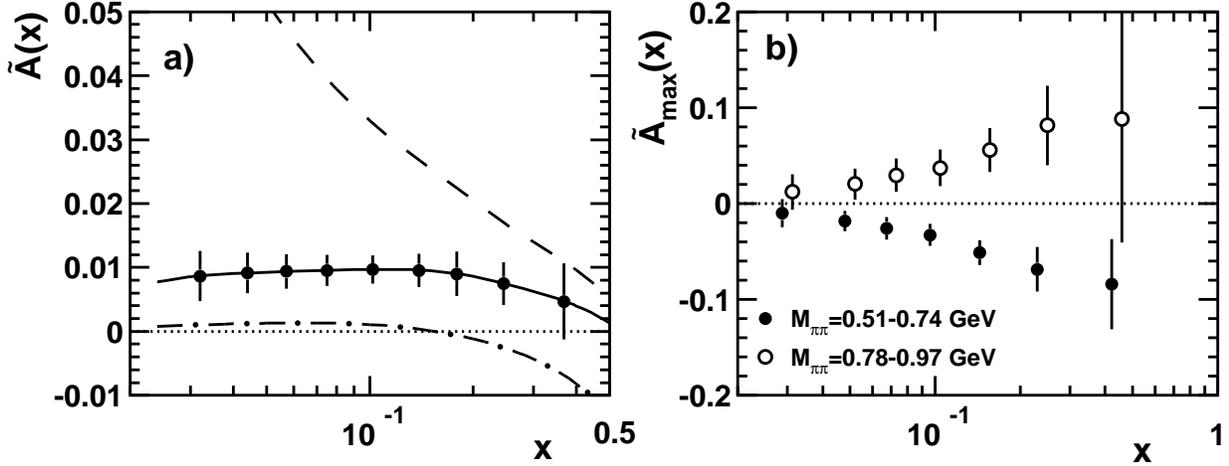,width=16.5cm} 
\caption{\it Projections for a measurement at HERMES of
a) twist-3 asymmetry for $\pi^+ + \pi^-$ production 
(dash-dotted curve --- $\delta q(x) =0 $, 
dashed curve --- saturation of the Soffer's inequality, 
solid curve --- $\delta q(x) = \Delta q(x)$), b) two-pion
asymmetry. }
\label{hermcomb}
\end{figure}

{\bf Twist-3 Pion Production.} An effect of the transversity distributions
in the spin-dependent cross-section of pion production in DIS of longitudinally
polarized leptons on a transversely polarized nucleon target can appear only
at the twist-3 level, when $\delta q(x)$ contributes through coupling 
with the chiral-odd twist-3 FF $\hat{e}( z )$ \cite{jaffe-ji93}.
A simple relation between $\hat{e}(z)$ and the unpolarized FF $D(z)$
has been predicted in the chiral quark model~\cite{jizhu}
$ \hat{e}(z) = z D(z) {m_q \over M} \approx {1 \over 3} z D(z) $,
where $m_q$ is the constituent quark mass.
In the particular case of forming the asymmetry for the sum of 
$\pi^+$ and $\pi^-$ production on a proton target:
\be
A(x,y,\phi) = \cos\phi \cdot \frac{2 M x}{\sqrt{Q^2}} \cdot
      \frac{2 y \sqrt{1 - y}}{1 + (1 - y)^2} \cdot
      { {g_T(x) + h_1(x)/3x - (1-y/2) g_1(x)} \over { F_1(x) } } ,
\ee
where $h_1(x) = \frac{1}{2} \sum_i e_i^2 \delta q_i(x)$, 
$g_T( x ) = g_1( x ) + g_2( x )$, and
$\phi$ is the azimuthal angle between the lepton scattering plane and the spin 
plane. The projections for a measurement of this twist-3 pion asymmetry
at HERMES are shown in fig.~\ref{hermcomb}a, where the asymmetry 
$\tilde{A}(x,y) = P_B \cdot P_T \cdot A(x,y,\phi) / \cos\phi$ is shown
for convenience.

{\bf Two-Meson Correlations with Transverse Spin.} 
As has been noted above, the asymmetry in this case depends on
the unknown chiral-odd interference quark FF $\delta \hat{q}_{_I}(z)$,
${\cal A}_{\bot\top} \sim 
 \sum_a e_a^2 \delta q_a(x)\, \delta \hat{q}_{_I}^a(z)$.

A {\it maximally} possible asymmetry with respect to the interference FF
can be obtained with its upper bound.
For $\pi^+ \pi^-$ pair production at a proton target the maximal asymmetry 
takes the form:
\be
A_{max}\, =\, 
-\frac{\pi}{\sqrt{32}} \, 
\cos{\phi}\,\sin\left(\delta_0-\delta_1\right)\, D_{nn}
{{\delta u_v(x)\,-\,\frac{1}{4}\delta d_v(x)} \over
 {(u(x) + \bar{u}(x))\,+\,\frac{1}{4}(d(x) + \bar{d}(x))}}\ ,
\label{twopiasy}
\ee
where $\cos\phi = {{\vec k_+}\times{\vec k_-}\cdot\vec S_\perp / |
\vec k_+\times \vec k_-||\vec S_\perp|}$;
$D_{nn}$ is the transverse polarization transfer coefficient,
and $\delta_{0,1} = \delta_{0,1}(m^2)$ are strong interaction $\pi\pi$ 
phase shifts.
The asymmetry has been 
calculated in two regions of the two-pion mass
to avoid averaging to zero due to
the factor $\sin\left(\delta_0-\delta_1\right)$. The projections
for its measurement at HERMES are shown in fig.~\ref{hermcomb}b in terms of
$\tilde{A}_{max}\,=\,P_T \cdot A_{max}/\cos\phi\ $.

\newpage

{\bf Collins effect.} In the case of an unpolarized beam 
and a transversely polarized target the following 
{\it weighted asymmetry}  \cite{kotmul97}
provides access to the quark transversity distribution via the Collins 
effect:
\be
A_T(x,y,z) \equiv
\frac{\int d \phi^\ell \int d^2P_{h\perp}\, 
\frac{\vert P_{h\perp}\vert}{zM_h}
\sin(\phi_s^\ell + \phi_h^\ell)
\,\left(d\sigma^{\uparrow}-d\sigma^{\downarrow}\right)}
{\int d \phi^\ell \int d^2 P_{h\perp} (d\sigma^{\uparrow}+d\sigma^{\downarrow})} ,
\label{collasy}
\ee
where $P_{h\perp}$ is the pion's transverse momentum and
the azimuthal angles are defined in the transverse
space giving the orientation of the lepton plane ($\phi^\ell$) and the
orientation of the hadron plane ($\phi^\ell_h$ = $\phi_h - \phi^\ell$) 
or spin vector ($\phi^\ell_s$ = $\phi_s - \phi^\ell$)
with respect to the lepton plane. 
The asymmetry (\ref{collasy}) can be estimated from
\be
\label{asymmetry}
A_T(x,y,z) =
P_T \cdot D_{nn} \cdot
\frac{\sum_q e^2_q \  \delta q(x) \ H_1^{\perp(1)q}(z)}
     {\sum_q e^2_q \  q(x) \ D^q_1(z)} .
\ee

The assumption of $u$-quark dominance was used to calculate 
the expected asymmetry $A_T^{\pi^+}(x)$~\cite{KNO}.
In this case the asymmetry  for a proton target reduces to 
\be
\label{uasymmetry}
A_T^{\pi^+}(x,y,z) =
P_T \cdot D_{nn} \cdot
\frac{\delta u(x)}{u(x)} \cdot \frac{H_1^{\perp(1)u}(z)}{D^u_1(z)} 
\ee
The approach of Ref.~\cite{kotmul97} is adopted to estimate
$H_1^{\perp(1)u}(z) / D^u_1(z)$. 

\begin{figure}[htb]
\centering
\epsfig{file=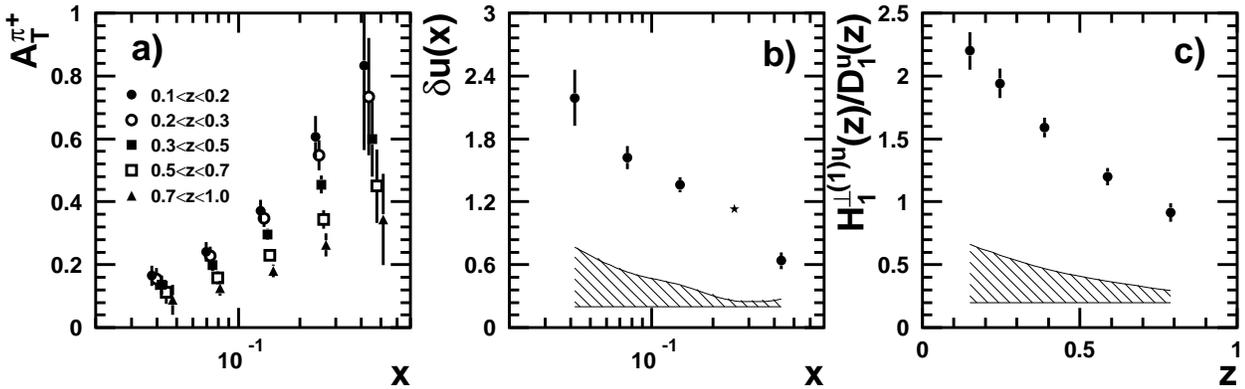,width=16.0cm}
\caption{\it a) The weighted asymmetry $A_T^{\pi^+}(x)$ in 
different intervals of $z$; b) the transversity distribution $\delta u(x)$, and
c) the ratio of the fragmentation functions 
$H_1^{\perp (1) u} (z) / D_1^u (z)$ 
as it would be measured by HERMES. 
The asterisk in b) shows the normalization point.  
The hatched bands in b) and c) show projected systematic uncertainties
due to the normalization and the $u$-quark dominance assumptions.}
\label{collins1}
\end{figure}

The factorized form of expression (\ref{uasymmetry}) with respect to 
$x$ and $z$ allows the simultaneous reconstruction
of the shape for both unknown functions $\delta u(x)$ and
$H_1^{\perp(1)u}(z) / D^u_1(z)$ if measurements of the asymmetry 
are done in $(x, z)$ bins, while the relative normalization
cannot be fixed without a further assumption. 
The differences between $\delta q(x)$ and $\Delta q(x)$ are smallest in 
the region of intermediate and large values of $x$ \cite{KNO}. 
Hence the assumption $ \delta q (x_0) = \Delta q(x_0) $
at $x_0=0.25$ was made to resolve the normalization ambiguity. 
The projections for a measurement of $A_T^{\pi^+}(x)$, $\delta u(x)$, and 
$H_1^{\perp(1)u}(z) / D^u_1(z)$ at HERMES 
are shown in fig.~\ref{collins1}.

\newpage

\subsection{The Experiment TESLA-N}

The basic TESLA-N~\cite{teslan} idea is to use one of 
the arms of the presently planned $e^+ e^-$ collider TESLA at DESY
to accomplish collisions of longitudinally polarized electrons ($E = 250$~GeV,
$P_B = 90\%$) with a polarized solid-state fixed target.
Presently, the target materials NH$_3$ ( $P_T$ = 0.8, f = 0.176)  
and $^6$LiD ($P_T$ = 0.3, f = 0.44) appear as the best choices to study 
electron scattering off polarized protons and deuterons, respectively.

The physics projections presented below are based on an integrated 
luminosity of 100 fb$^{-1}$. This represents a conservative estimate for 
{\it one} year of data taking.

Measurements of single target-spin asymmetries due to the Collins effect 
in the production 
of positive and negative pions on proton and deuteron targets 
($A_{p, d}^{\pi^+, \pi^-}$)
allow, under reasonable assumptions, the simultaneous reconstruction
of the shapes of the unknown functions $\delta q(x,Q^2)$ and
$H_1^{\perp(1)}(z) / D_1(z)$. Again the relative normalization
cannot be fixed without an independent measurement or a further assumption. 
This ambiguity can be resolved by relating $\delta q(x)$ to $\Delta q(x)$
at small values of $Q^2$.
Following ref.~\cite{KNO}, the
normalization ambiguity is resolved by assuming
$\delta u (x_0, Q_0^2) = \Delta u(x_0, Q_0^2)$ at $x_0 = 0.25$.

\begin{figure}[htb]
\centering
\epsfig{file=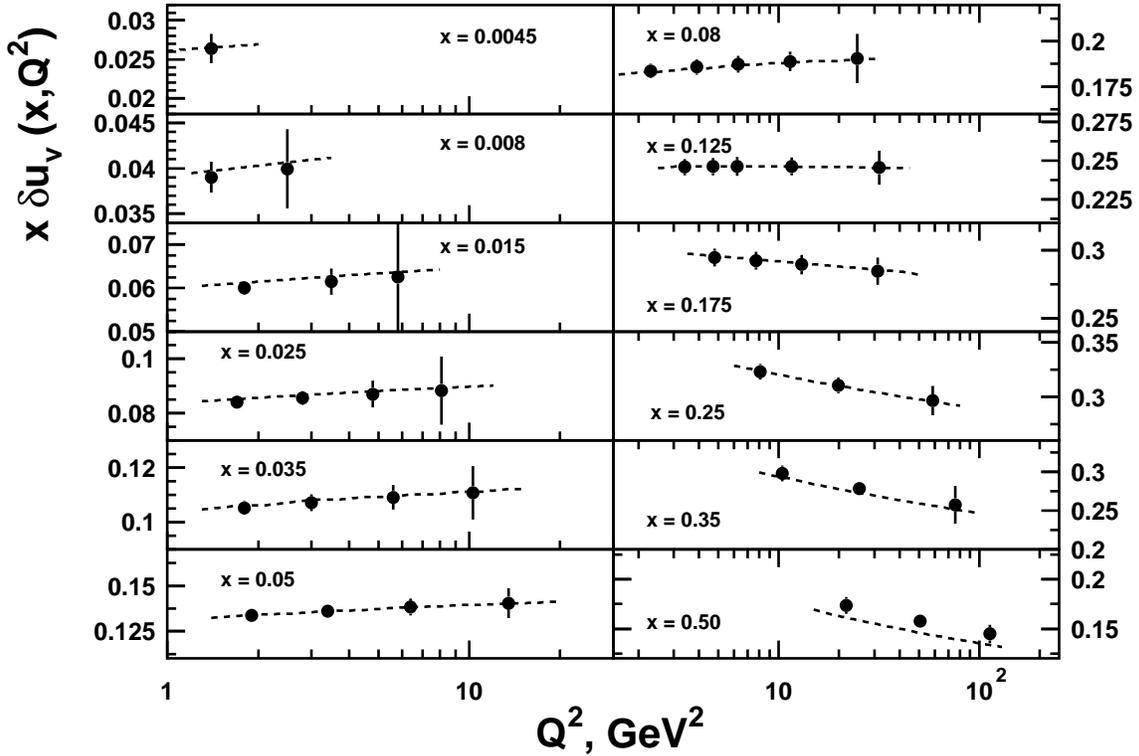,width=15.0cm}
\caption{\sf The transversity distribution $\delta u_v( x, Q^2)$ 
as it would be measured at TESLA-N. The curves show the LO
$Q^2$-evolution of $\delta u_v$ obtained with a fit to the simulated 
asymmetries.}
\label{figure:h1uval}
\end{figure}

The projections for the measurement of 
$\delta u_v (x, Q^2)$ at TESLA-N are shown in fig.~\ref{figure:h1uval}.
A broad range of $0.003 < x <0.7$ can be accessed
in conjunction with $1 < Q^2 < 100$ GeV$^2$.
A simultaneous reconstruction of the quark transversity distributions  
$\delta d_v$, $\delta \bar{u}$, and $\delta \bar{d}$ 
is attained with a somewhat lower accuracy. 
Projections for the accuracies of a measurement 
of the $u$- and $d$-quark tensor charges are
$\pm 0.01$ and $\pm 0.02$ at the scale of 1 GeV$^2$, respectively. 
At the same time, precise values would be measured for the ratio of 
polarized and unpolarized favored quark fragmentation functions 
$H_1^{\perp(1)}(z)/D_1(z)$, assumed to be flavor-independent
in this analysis.

\newpage

\section{Summary}

A measurement of the transversity distributions in polarized proton-proton
collisions at RHIC will be possible if the chiral-odd interference fragmentation
function, $\delta\hat{q}_{_I}$, 
is not heavily suppressed relative to its theoretical upper bound.

The HERMES experiment using a transversely polarized proton target
will be capable in a few years to 
measure simultaneously and with good statistical precision the 
u-quark transversity distribution $\delta u(x)$ and the 
ratio of the fragmentation functions $H_1^{\perp (1) u}(z) / D_1^u (z)$.

A measurement of the quark transversity distributions 
as a function of $x$ and $Q^2$
with good statistical accuracy requires a new high luminosity and high
energy polarized lepton-nucleon experiment.
The physics potential of the TESLA-N project
is demonstrated by showing that
an accurate measurement of the $x$- and $Q^2$-dependence of
the transversity quark distributions would be possible.

%
%

\end {document}